\documentclass[12pt]{iopart}
\usepackage{graphicx}% Include figure files
\usepackage{bm}% bold math
\usepackage{color}
\usepackage[utf8]{inputenc}
\usepackage[T1]{fontenc}
\usepackage{mathptmx}
\usepackage{wrapfig, blindtext}
\usepackage{float} % needed for figures
\usepackage{subfigure}
\usepackage{amssymb}   % for math
\usepackage{siunitx}   % for si units such as angstrom used as \SI{1}{\angstrom}
\usepackage{gensymb} %degree symbol \degree
\usepackage{textcomp} %for angle brackets in text mod \textlangle{} and for commands like \textminus
\usepackage{csquotes} %for comma
\usepackage{soul}
\usepackage[colorlinks=true, urlcolor=blue, citecolor=blue, linkcolor=blue]{hyperref}
\setstcolor{red}
\usepackage{setspace}
\newcommand{\im}{\mathrm{i}}
\usepackage[dvipsnames]{xcolor}
\usepackage{wrapfig}

\usepackage[colorlinks=true, urlcolor=blue, citecolor=blue, linkcolor=blue]{hyperref}
\begin{document}
\title{Quantum information diode based on a magnonic crystal}
\author{Rohit K. Shukla}
\address{ Department of Physics, Indian Institute of Technology (Banaras Hindu University) Varanasi - 221005, India}
\author{Levan Chotorlishvili}
\ead{levan.chotorlishvili@gmail.com}
\address{ Department of Physics and Medical Engineering, Rzeszow University of Technology, 35-959 Rzeszow Poland}
\author {Vipin Vijayan}
\address{ Department of Physics, Indian Institute of Technology (Banaras Hindu University) Varanasi - 221005, India}
\author{Harshit Verma}
\address{ Centre for Engineered Quantum Systems (EQUS), School of Mathematics and
Physics, The University of Queensland, St Lucia, QLD 4072, Australia}
\author{Arthur Ernst}
\address{Max Planck Institute of Microstructure Physics, Weinberg 2, D-06120 Halle, Germany}
\address{Institute of Theoretical Physics, Johannes Kepler University Alterger Strasse 69, 4040 Linz, Austria}
\ead{Arthur.Ernst@jku.at}
\author{Stuart S.\,P. Parkin}
\address{Max Planck Institute of Microstructure Physics, Weinberg 2, D-06120 Halle, Germany}
\author{Sunil K. Mishra}
\ead{sunilkm.app@iitbhu.ac.in}
\address{ Department of Physics, Indian Institute of Technology (Banaras Hindu University) Varanasi - 221005, India}

\begin{abstract}
Exploiting the effect of nonreciprocal magnons in a system with no
inversion symmetry, we propose a concept of a quantum information
diode, {\it i.e.}, a device rectifying the amount of quantum
information transmitted in the opposite directions. We control the
asymmetric left and right quantum information currents through an
applied external electric field and quantify it through the left and
right out-of-time-ordered correlation (OTOC).  To enhance the efficiency of the quantum information diode, we utilize a magnonic crystal. We excite magnons of different frequencies and let them
propagate in opposite directions. Nonreciprocal magnons propagating in
opposite directions have different dispersion relations. Magnons
propagating in one direction match resonant conditions and scatter on
gate magnons. Therefore, magnon flux in one direction is damped in the
magnonic crystal leading to an asymmetric transport of
quantum information in the quantum information diode. A quantum
information diode can be fabricated from an yttrium iron garnet (YIG)
film. This is an experimentally feasible concept and implies certain
conditions: low temperature and small deviation from the equilibrium
to exclude effects of phonons and magnon interactions. We show that
rectification of the flaw of quantum information can be controlled
efficiently by an external electric field and magnetoelectric
effects. 
\end{abstract}

\maketitle
\section{Introduction}
A diode is a device designated to support asymmetric transport.
Nowadays, household electric appliances or advanced experimental
scientific equipment are all inconceivable without extensive use of
diodes. Diodes with a perfect rectification effect permit electrical
current to flow in one direction only.  The progress in nanotechnology
and material science passes new demands to a new generation of diodes;
futuristic nano-devices that can rectify either acoustic (sound
waves), thermal phononic, or magnonic spin current transport.
Nevertheless, we note that at the nano-scale, the rectification effect is
never perfect, i.e., backflow is permitted, but amplitudes of the
front and backflows are different
\cite{Liang,ZhangCheng,Chen,Maldovan,Ren,Lepri,
  Komatsu,Majumdar,Terasaki,Wang,Casati,WangCasati,LiRen, Etesami}. In
the present work, we propose an entirely new type of diode designed to
rectify the quantum information current. We do believe that in the
foreseeable future the quantum information diode (QID) has a perspective
to become a benchmark of quantum information technologies.  

The functionality of a QID relies on the use of magnonic crystals,
{\it i. e.}, artificial media with a characteristic periodic lateral
variation of magnetic properties. Similar to photonic crystals,
magnonic crystals possess a band gap in the magnonic excitation
spectrum. Therefore, spin waves with frequencies matching the band gap
are not allowed to propagate through the magnonic crystals
\cite{Chumak1,Chumak2,Nikitov,Kruglyak,ZKWang,Gubbiotti,ABUstinov}. This effect has been utilized earlier to demonstrate a magnonic transistor in a YIG strip \cite{Chumak1, Chumak2}.   
The essence of a magnonic transistor is a YIG strip with a periodic
modulation of its thickness (magnonic crystal). The transistor is
complemented by a source, a drain, and gate antennas. A gate antenna
injects magnonic crystal magnons with a frequency $\omega_G$ matching
the magnonic crystal band gap. In the process, the gate magnons cannot leave
the crystal and may reach a high density. Magnons emitted from a
source with a wave vector $ \textbf{k}_s$ flowing towards the drain run
into the magnonic crystal. The interaction between the source magnons and
the magnonic crystal magnons is a four-magnon scattering process. The
magnonic current emitted from the source attenuates in the magnonic
crystal, and the weak signal reaches the drain due to the scattering.
The relaxation process is swift if the following condition holds
\cite{Chumak2,Gurevich} 
\begin{equation}
\label{certain conditions} 
  k_s=\frac{m_0\pi}{a_0}, 
\end{equation} 
where $m_0$ is the integer,
and $a_0$ is the crystal lattice constant. The magnons with wave
vectors satisfying the Bragg conditions Eq.~(\ref{certain conditions})
will be resonantly scattered back, resulting in the generation of
rejection bands in a spin-wave spectrum over which magnon propagation
is entirely prohibited. Experimental verification of this effect is
given in Ref.~\cite{Chumak2}. 
\begin{figure}[t!]
  \centering
 \includegraphics[width=.75\linewidth, height=.4\linewidth]{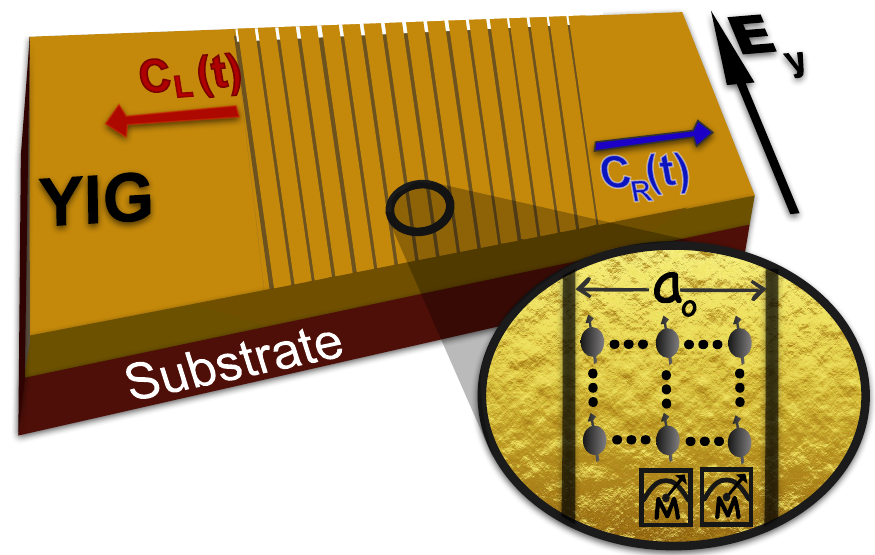}
  \caption{Illustration of a quantum information diode: A plane of a YIG film with grooves orthogonal to the direction of the
    propagation of quantum information. In the middle of the QID, we
    pump extra magnons to excite the system. A quantum excitation
    propagates toward the left, and the right ends asymmetrically. To
    describe the propagation process of quantum information, we
    introduce the left and right OTOC $C_L(t)$ and $C_R(t)$. Because
    the left-right inversion is equivalent to $D\rightarrow-D$, meaning
    $E_y\rightarrow-E_y$, we can invert the left and right OTOCs by
    switching the applied external electric field.}
  \label{QI_DIODE}
\end{figure}
\section{Results}
\subsection{ Proposed set-up for QID}
A pictorial representation of a QID is shown in Fig.~\ref{QI_DIODE}. A
magnonic crystal can be fabricated from a YIG film. Grooves can be
deposited using a lithography procedure in a few nanometer steps, and,
for our purpose, we consider parallel lines in width of 1$\mu$m spaced
with 10$\mu$m from each other. Therefore, the lattice constant,
approximately $a_0=11\mu$m, {\it i. e.}, is much larger than the unit
cell size $a=10$nm used in our coarse-graining approach. Due to the
capacity of our analytical calculations, we consider quantum spin
chains of length about $N=1000$ spins and the maximal distance between
the spins $r_{ij}=d$ (in the units of $a$), $d=i-j=40$. In what
follows, we take $k(\omega)a\ll1$. The mechanism of the QID is based
on the effect of direction dependence of nonreciprocal magnons
\cite{Takashima,Matsumoto,Shiomi}. In the chiral spin systems, the
absence of inversion symmetry causes a difference in dispersion
relations of the left and right propagating magnons, {\it i. e.,}
$\omega_{s,L}(\textbf{k})\neq\omega_{s,R}(-\textbf{k})$. Due to the
Dzyaloshinskii–Moriya interaction (DMI), magnons of the same frequency
$\omega_s$ propagating in opposite directions have different wave
vectors \cite{LevanWang}: $a\left(k^+_s-k^-_s\right)=D/J$, where $J$
is the exchange constant, and $D$ is the DMI constant. Therefore, if
the condition Eq.(\ref{certain conditions}) holds for the left
propagating magnons, it is violated for the right propagating magnons
and vice versa. These magnons propagating in different directions
decay differently in the magnonic crystal. Without loss of generality,
we assume that the right propagating magnons with $k^+_s$ satisfy the
condition Eq.(\ref{certain conditions}), and the current attenuates
due to the scattering of source magnons by the gate magnons. The left
propagating magnons $k^-_s$ violate the condition Eq.(\ref{certain
  conditions}), and the current flows without scattering. Thus,
reversing the source and drain antennas' positions rectifies the
current.  Following ref.~\cite{Chumak2}, we introduce a suppression rate of the source to drain the magnonic current $\xi(D)=1-n_D^+/n_D^-$, where $n_D^+<n_D^-$ are densities of the drain magnons with and without scattering. The parameter $\xi(D)$ is experimentally accessible, and it depends on a particular setup. Therefore, in this manuscript, we take $\xi(D)$ as a free theory parameter. Multiferroic (MF) materials are considered as a good
example of a system with broken inversion symmetry, (see
Refs.~\cite{Nagaosa,Mostovoy,Ramesh,Bibes,Fiebig,Hemberger,Meyerheim,Cheong,Vedmedenko})
and references therein. MF properties of YIG are studied in
ref.~\cite{Vignale}. Moreover, in accordance with the scanning
tunneling microscopy experiments, a change in the spin direction at
one edge of a chiral chain was experimentally probed by tens of
nanometers away from the second edge \cite{Vedmedenko}.
\subsection{Model}
We consider a
2D square-lattice spin system with nearest-neighbor $J_1$ and the next
nearest-neighbor $J_2$ coupling constants:
\begin{equation}
\label{Hamiltonian}
\hat{H}=J_1\sum\limits_{\langle
 n,m\rangle}\hat\sigma_n\hat\sigma_m+
  J_2\sum\limits_{\langle\langle
  n,m\rangle\rangle}\hat\sigma_n\hat\sigma_m-{\bf
  P}\cdot{\bf E}, 
\end{equation} 
where $\langle n,m\rangle$
and $\langle\langle n,m\rangle\rangle$ indicates all the pairs with nearest-neighbor and next nearest-neighbor interactions, respectively. The last term in Eq.~(\ref{Hamiltonian}) describes a coupling of the ferroelectric polarization with unit vector $\mathbf{e}^x_{i,i+1}$, 
$\mathbf{P}=g^{\phantom{\dagger}}_{\mathrm{ME}}\mathbf{e}^x_{i,i+1}\times\left(\hat\sigma_i\times\hat\sigma_{i+1}\right)$
with an applied external electric field and mimics an
effective Dzyaloshinskii–Moriya interaction term
$D=E_y g^{\phantom{\dagger}}_{\mathrm{ME}}$ breaking the
left-right symmetry, where
\begin{equation}
-{\bf P}\cdot{\bf E}=D\sum\limits_{n}(\hat\sigma_{n}\times\hat\sigma_{n+1})_z.
\end{equation} 
Here we consider only the nearest neighbor
DMI and only in one direction. As a consequence, the
left-right inversion is equivalent to $D\rightarrow-D$, or
$E_y\rightarrow-E_y$. The broken left-right inversion
symmetry can be exploited in rectifying the information
current by an electric field. More importantly, the procedure is
experimentally feasible. We can diagonlize the Hamiltonian in Eq.~(\ref{Hamiltonian})
%utilize the following rotation:
%$\hat\sigma_n^{\alpha}\rightarrow\gamma_{n}^{\alpha}\hat\sigma_n^{z}+(A_{n}^{\alpha}+A_{n}^{\ast\alpha})\hat\sigma_n^{x}\pm
%i(A_{n}^{\alpha}-A_{n}^{\ast\alpha})\hat\sigma_n^{y}$,
%where $\alpha=x,y,z$. The coefficients of the rotation are
%given by:
%$A_{n}^{x}=-e^{i\varphi_{n}}(1+\gamma_{n}^{z})/4+e^{-i\varphi_{n}}(1-\gamma_{n}^{z})/4$,
%$A_{n}^{y}=i\big(e^{i\varphi_{n}}(1+\gamma_{n}^{z})/4+e^{-i\varphi_{n}}(1-\gamma_{n}^{z})/4\big)$,
%$A_{n}^{z}=\sqrt{1-(\gamma_{n}^{z})^{2}}/2$,
%$\gamma_{n}^{x}=\sqrt{1-(\gamma_{n}^{z})^{2}}\cos\varphi_{n},
%\gamma_{n}^{y}=\sqrt{1-(\gamma_{n}^{z})^{2}}\sin\varphi_{n}$.
by using the Holstein-Primakoff transformation
\cite{FNori,Udvardi,Zheng,Stagraczynski}[See \ref{supp0}
for detailed
derivation] % Eq.(\ref{Hamiltonian}) can be written in the form of bosonic creation and anihilation operators
as: %\begin{widetext}
  \begin{eqnarray}\label{2Hamiltonian}
  &&  \hat{H} = \sum\limits_{\vec{k}}\omega(\pm D,\textbf{k})\hat{a}^{\dagger}_{\vec{k}}\hat{a}_{\vec{k}},\,\,\omega(\pm D,\textbf{k})=\big(\omega(\vec{k})\pm\omega_{DM}(\vec{k})\big),  ~~\omega_{DM}(\vec{k})=D\sin(k_xa),\nonumber\\
 &&\omega(\vec{k})=2J_1(1-\gamma_{1,\textbf{k}})+2J_2(1-\gamma_{2,\textbf{k}}),\,\,gamma_{1,\textbf{k}}=\frac{1}{2}(\cos k_xa+\cos k_ya),\nonumber\\
 && \gamma_{2,\textbf{k}}=\frac{1}{2}[\cos (k_x+k_y)a+\cos (k_x-k_y)a].
  \end{eqnarray}
%\end{widetext}
Here $\pm D$ corresponds to the magnons propagating in opposite
directions and the sign change is equivalent to the electric field
direction change.  We note that a 1D character of the DM term is
ensured by the magnetoelectric effect \cite{Nagaosa} and to the electric field applied along the $\textbf{y}$ axis. 

 The speed limit of information propagation is usually given in terms of Lieb-Robinson (LR) bound, defined for the Hamiltonians that are  locally bounded and short-range interacting \cite{kuwahara2021lieb,hastings2006spectral,nachtergaele2006propagation}. Since the Hamiltonian in Eq.~(\ref{Hamiltonian}) satisfies both conditions, the LR bound can be defined for the spin model. However, when we transform the Hamiltonian using Holstein-Primakoff bosons, we have to take extra care as the bosons are not locally bounded. To define LR bound, we take only a few noninteracting magnons and exclude the magnon-magnon interaction to truncate terms beyond quadratic operators. In a realistic experimental setting, a low density of propagating magnons in YIG can easily be achieved by properly controlling the microwave antenna. In the case of low magnon density, the role of the magnon-magnon interaction between propagating magnons in YIG is negligible. Therefore, for YIG, we have a quadratic Hamiltonian, which is a precise approach in a low magnon density limit. Our discussion is valid for the experimental physical system \cite{Chumak2}, where magnons of YIG do not interact with each other, implying that there is no term in the Hamiltonian beyond quadratic. We can estimate LR bounds \cite{LiebRobinson} defining the maximum group velocities of the left-right propagating magnons $v_{g}^{\pm}(\vec{k})=\frac{\partial (\omega(\vec{k})\pm\omega_{DM}(\vec{k}))}{\partial k}$. Taking into account the explicit form of the dispersion relations, we see that the maximal asymmetry is approximately equal to the DM constant {\it i. e.,} $v_{g}^{+}(0)-v_{g}^{-}(0)\approx 2D$. We note that the effect of nonreciprocal magnons is already observed experimentally {\cite{KZakeri, IguchiUemura, ShibataKubota, TaguchiArima, Gitgeatpong}} but up to date, never discussed in the context of the quantum information theory.

% We would like to study the propagation of a single spin-flip on
% ferromagnetic ground state and estimate Lieb–Robinson (LR) bounds as
% discussed in
% Ref.~\cite{Zheng} %for spin-$1/2$ model with only one magnon from a
% fully polarized state for which mapping to a quadratic bosonic
% Hamiltonian is exact. 
% {\color{blue} For LR velocity, one can consider the whole Hilbert
% space. However, this is possible for finite toy models of a small
% number of spins. Such simplified toy models and full basis
% calculations for the LR velocity are not experimentally relevant and
% interesting for a broader audience. In the present work, we proposed
% an experimentally feasible model and approximation we exploited when
% calculated LR is physically motivated and relevant to a particular
% setup. LR bounds defining the maximum group velocities of the
% left-right propagating magnons
% $v_{g}^{\pm}(\vec{k})=\frac{\partial
% (\omega(\vec{k})\pm\omega_{DM}(\vec{k}))}{\partial k}$. Taking into
% account the explicit form of the dispersion relations we see that
% the maximal asymmetry is approximately equal to the DM constant {\it
% i. e.,} $v_{g}^{+}(0)-v_{g}^{-}(0)\approx 2D$. We note that the
% effect of nonreciprocal magnons is already observed experimentally
% \cite{KZakeri, IguchiUemura, ShibataKubota, TaguchiArima,
% Gitgeatpong} but up to date, never discussed in the context of the
% quantum information theory.  }
We formulate the central interest question as follows: At $t=0$, we
act upon the spin $\hat\sigma_{n}$ to see how swiftly changes in the
spin direction %of the spin $\hat\sigma_{n}$%
can be probed tens of sites away $d=n-m\gg 1$, and whether the forward
and backward processes (i.e., probing for $\hat\sigma_{m}$ the outcome
of the measurement done on $\hat\sigma_{n}$) are asymmetric or not.
Due to the left-right asymmetry, the chiral spin channel may sustain a
diode rectification effect when transferring the quantum information
from left to right and in the opposite direction. We note that our
discussion about the left-right asymmetry of the quantum information
flow is valid until the current reaches boundaries. Thus the upper
limit of the time reads $t_{max}=Na/v_{g}^{\pm}(\vec{k})$, where $N$
is the size of the system.

\subsection{Out-of-time-order correlator}
Larkin and Ovchinnikov \cite{Larkin} introduced the concept of the out-of-time-ordered correlator (OTOC), and since then, OTOC has been seen as a diagnostic tool of quantum chaos. The concern of
delocalizations in the quantum information theory (i.e., the scrambling of quantum
entanglement) was renewed only recently, see Refs.~\cite{Maldacena, Roberts, Iyoda, Chapman, Swingle, Schmalian, delCampo, Campisi, Hosur, Halpern} and references therein. OTOC is also used for describing the static and dynamical phase transitions  \cite{heyl2018detecting,chen2020detecting, shukla2021}. Dynamics of the semi-classical, quantum, and spin systems can be discussed by using OTOC  \cite{rozenbaum2017lyapunov, Maldacena,shukla2022out,shukla2022characteristic, kukuljan2017weak}. We utilize OTOC to characterize the left-right asymmetry of the quantum information flow and thus infer the rectification effect of a diode. 

Let us consider two unitary operators $\hat{V}$ and $\hat{W}$
describing local perturbations to the chiral spin system
Eq.~(\ref{Hamiltonian}), and the unitary time evolution of one of the
operators
${\hat{W}\left(t\right)=\exp(i\hat{H}t)\hat{W}(0)\exp(-i\hat{H}t)}$.
Then the OTOC is defined as \begin{equation}\label{OTOC1}
  C\left(t\right)=
  \frac{1}{2}\left\langle\left[\hat{W}(t),\hat{V}(0)\right]^{\dag}\left[\hat{W}(t),\hat{V}(0)\right]\right\rangle,
\end{equation} where parentheses $\langle \cdots\rangle$ denotes a
quantum mechanical average over the propagated quantum state. Following
the definition the OTOC at the initial moment is 
zero ${C(0)=0}$, provided that ${[\hat W(0),\hat V(0)]=0}$. In
particular, for the local unitary and Hermitian operators of our choice
$\hat W_m^{\dagger}\left(t\right)\equiv\hat \sigma_m^z(t)=\exp(i\hat
Ht) \hat \eta_m\exp(-i\hat Ht)$,
and $\hat V_n^\dagger=\hat \sigma_n^z=\hat \eta_n$, where
$\hat \eta_n=\hat 2a^\dagger_n\hat{a}_n-1$. The bosonic operators are
related to the spin operators via $\sigma_n^-=2a_n^{\dagger}, \ \ \sigma_n^+=2a_n, \ \
  \sigma_n^z=2a_n^{\dagger}a_n-1$. In terms of the
occupation number operators, the OTOC is given as
\begin{eqnarray}
\label{occupation number}
  C(t)&=&\frac{1}{2}\bigg\lbrace\langle\eta_n\eta_m(t)\eta_m(t)\eta_n\rangle+\langle \eta_m(t)\eta_n\eta_n\eta_m(t)\rangle \nonumber \\&-&\langle \eta_m(t)\eta_n\eta_m(t)\eta_n\rangle-\langle
          \eta_n\eta_m(t)\eta_n\eta_m(t)\bigg\rbrace. 
\end{eqnarray}
Indeed, the OTOC can be interpreted as the overlap of two wave
functions, which are time evolved in two different ways for the same
initial state $|\psi(0)\rangle$. The first wave function is obtained
by perturbing the initial state at ${t=0}$ with a local unitary
operator $\hat{V}$, then evolved further under the unitary evolution
operator ${\hat{U}=\exp (-\im\hat{H}t)}$ until time $t$. It is then
perturbed at time $t$ with a local unitary operator $\hat{W}$, and
evolved backwards from $t$ to ${t=0}$ under
$\hat{U^{\dagger}}$. Hence, the time-evolved wave function is
$|\psi(t)\rangle
=\hat{U^{\dagger}}\hat{W}\hat{U}\hat{V}|\psi(0)\rangle=\hat{W}(t)\hat{V}|\psi(0)\rangle$. To
get the second wave function, the order of the applied perturbations
is permuted, {\it i. e.,}~first $\hat{W}$ at $t$ and then $\hat{V}$ at
${t=0}$. Therefore, the second wave function is
$|\phi(t)\rangle=\hat{V}\hat{U^{\dagger}}\hat{W}\hat{U}|\psi(0)\rangle=\hat{V}\hat{W}(t)|\psi(0)\rangle$
and their overlap is equivalent to
$F(t)=\langle\phi(t)|\psi(t)\rangle$. The OTOC is calculated from this
overlap using $C(t)=1-\Re[{F(t)}]$. 
What breaks the time inversion symmetry for the OTOC is the permuted
sequence of operators $\hat{W}$ and $\hat{V}$. However, in
spin-lattice models with a preserved spatial inversion symmetry
$\mathcal{\hat{P}}\hat{H}=\hat{H}$, the spatial inversion
${\mathcal{\hat{P}}d(\hat{W}, \hat{V})}={-d(\hat{W},
  \hat{V})}={d(\hat{V}, \hat{W})}$ can restore the permuted order
between $\hat{V}$ and $\hat{W}$,  where $d(\hat{W}, \hat{V})$ denotes
the distance between observables $\hat{W}$ and  $\hat{V} $. Permuting
just a single wave function, one finds
$C(t)=1-\Re(\langle\phi(t)|\mathcal{\hat{P}}\mathcal{\hat{T}}|\psi(t)\rangle)=C(0)$. Thus,
a scrambled quantum entanglement formally can be unscrambled by a
spatial inversion. However, in chiral systems
${\mathcal{\hat{P}}\hat{H}\neq\hat{H}}$ and the unscrambling procedure
fails. 
\begin{figure}
\includegraphics[width=.33\linewidth, height=.25\linewidth] {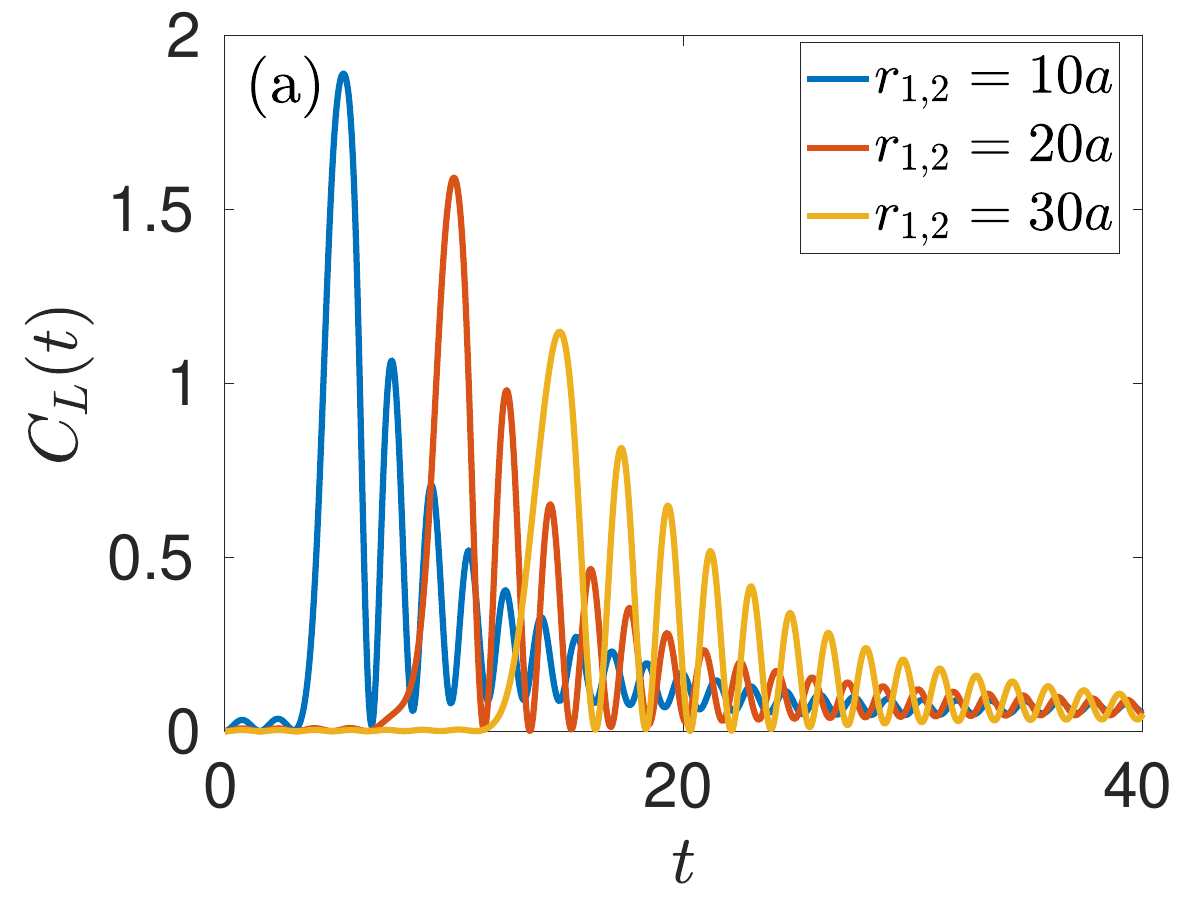}
 \includegraphics[width=.33\linewidth, height=.25\linewidth] {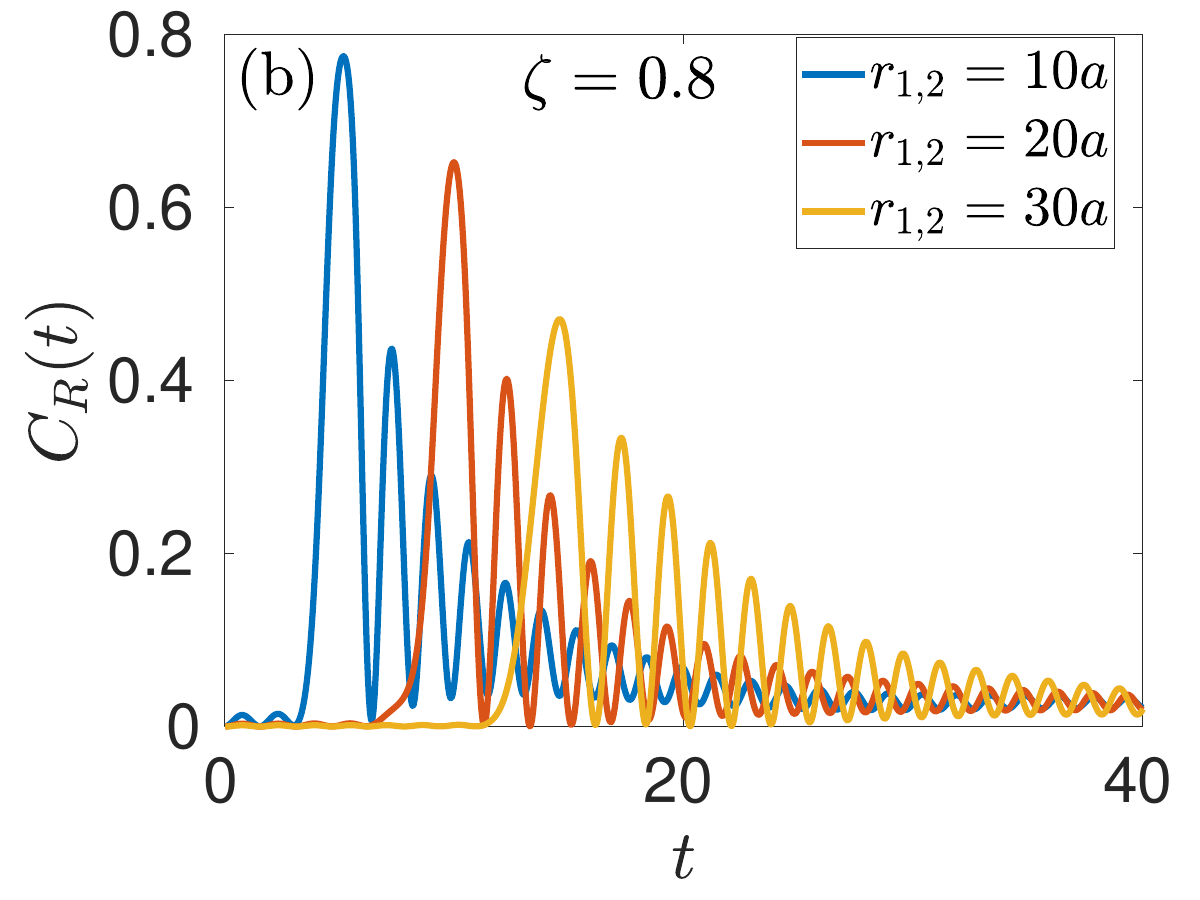}
  \includegraphics[width=.33\linewidth, height=.25\linewidth]{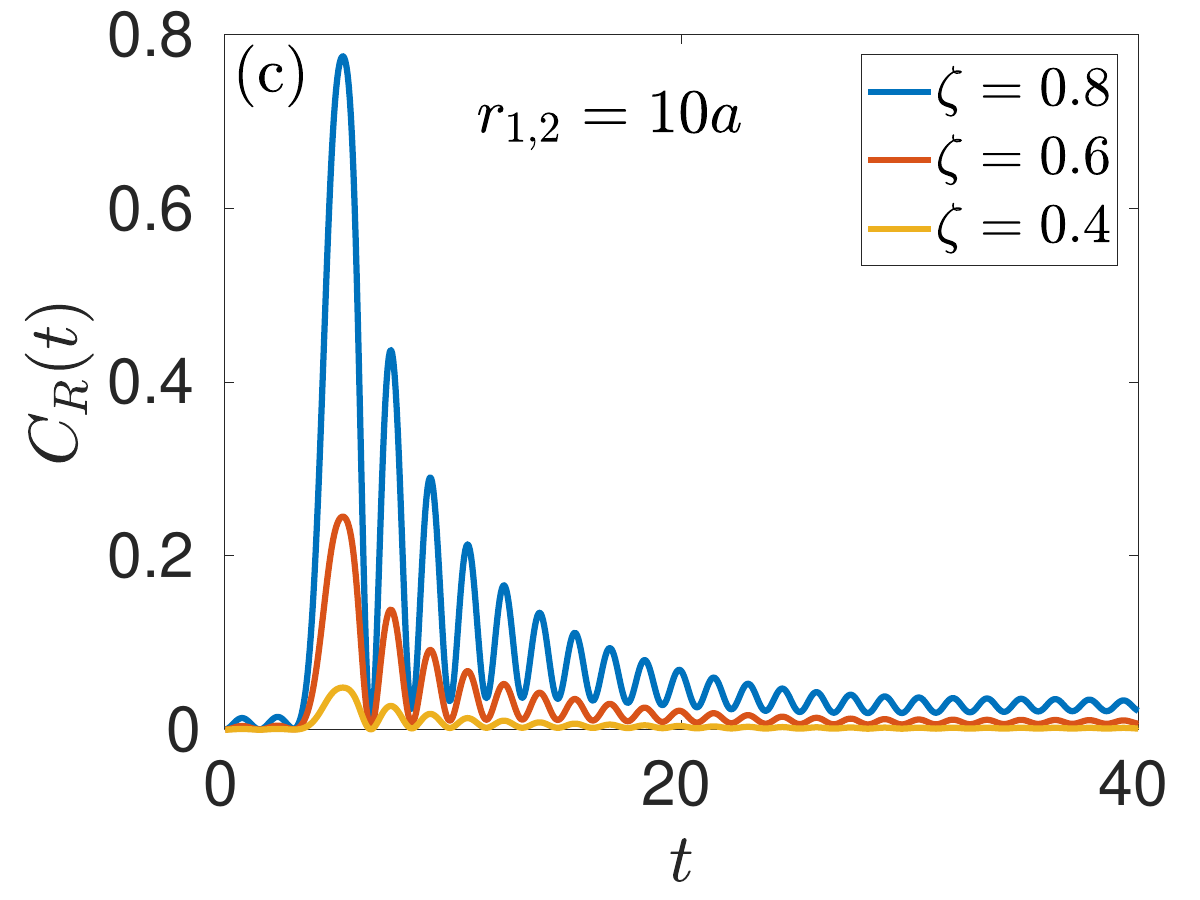}
\caption{\textbf{(a)} Left-OTOC and \textbf{(b)} Right-OTOC with time $t$ (in the units of $1/J$) for different distances $r_{1,2}=10a$, $20a$ and $30a$. \textbf{(c)} Right-OTOC with time for $r_{1,2}=10a$ and suppression rates of the magnon current $\zeta=0.8$, $0.6$ and $0.4$. Parameters are $N=1000$, $D=J_1=2J_2=1$. Periodic boundary conditions are considered. The values of the parameters: $m_0=1$ to $N$, $a=10^{-3}$ and $a_0=1$.}
\label{QI_DIODE_OTOC}
\end{figure}

\par
%{\color{blue} OTOC is directly related to second Renyi entropy ($S_V^2$), a measure of entanglement and  this relation is named as OTOC-RE theorem at infinite temperature \cite{Hosur2016,FAN}. In the mathematical form, it is given as $ C(n)\sim 1-e^{-S_V^2}= 1-\Tr \rho_V^2$,  where $S_V^2=-\ln \Tr_V(\rho_V^2)$ \cite{FAN,Bergamasco} and $\rho_V=\Tr_W[\rho]$ is the reduced density matrix.}
Taking into account Eq.~(\ref{2Hamiltonian}), we analyze quantum
information scrambling along the $\textbf{x}$ axis {\it i. e.}, 
$\omega(\pm D,\textbf{k})=\omega(\pm D, k_x, 0)$ and along
the $\textbf{y}$ axis, $\omega(0,\textbf{k})=\omega(0, 0, k_y)$. It is easy
to see that the quantum information flow along the $\textbf{y}$ axis is
symmetric, while along the $\textbf{x}$ axis it is
asymmetric and depends on the sign of the DM constant, i.e., the flow along
the $\textbf{x}$ is different from $-\textbf{x}$. Let us assume that
Eq.~(\ref{QI_DIODE}) holds for right-moving magnons 
and is violated for left-moving magnons. Excited magnons with the same
frequency and propagating into different directions have different
wave vectors 
$\omega_s\left(D,k_s^+\right)=\omega_s\left(-D,k_s^-\right)$ where: 
\begin{equation}
    \label{propagated wave modes}
\omega_s\left(\pm D,k_s^{\pm}\right)=2J_1(1-1/2\cos k^{\pm}_xa)+2J_2(1-\cos k^{\pm}_xa) \pm D\sin k^{\pm}_xa,
\end{equation}
$k^+_{m_0x}=\frac{m_0\pi}{a_0}$, $m_0=\mathbb N$ and $k^-_{m_0x}$ we
find from the condition
$\omega_s\left(D,k_s^+\right)=\omega_s\left(-D,k_s^-\right)$ leading
to
$k^-_{m_0x}=k^+_{m_0x}+\frac{2}{a}\tan^{-1}\left(\frac{D}{J_1+2J_2}\right)$.
Here we use shortened notations
$\omega_{m_0}=\omega_s\left(D,k_s^+\right)=\omega_s\left(-D,k_s^-\right)$
and set dimensionless units $J_1=2J_2\equiv J=1$. We excite in the diode magnons of different frequencies $m_0=[1,\,N]$.  Considering Eq.~(\ref{occupation number}), Eq.~(\ref{propagated wave modes}) and following Ref.~\cite{Zheng}, we obtain expressions for the left and right OTOCs $C_L(t)$ and $C_R(t)$ as: 
\begin{eqnarray}
%\label{we deduce the analytical formula2}
C_L(t)&=&\frac{8}{N^2}\Omega_1^L\Omega_2^L-\frac{8}{N^4}\Omega_1^L\Omega_2^L\Omega_1^L\Omega_2^L,  \nonumber \\
C_R(t)&=&\zeta^4(D)\Big(\frac{8}{N^2}\Omega_1^R\Omega_2^R-\frac{8}{N^4}\Omega_1^R\Omega_2^R\Omega_1^R\Omega_2^R\Big),  
\end{eqnarray}
where frequencies $\Omega_{1/2}^{L/R}$ and details of derivations are presented in \ref{supp1}.
Parameter $\xi$  enters into the right OTOC $C_R(t)$ expression because the right propagating magnons are scattered on the gate magnons. This is due to the non-reciprocal magnon dispersion relations associated with the DMI term.  Since the value of  $\xi$ depends on the experimental setup, we consider experimentally feasible values in our calculations.
\par
It should be noted that in the calculation of OTOC, we consider expectation value over the one magnon excitation state $\hat a^{\dagger}_n\vert \phi\rangle$, where $\vert \phi\rangle$ is the vacuum state. Such a state shows the presence of the quantum blockade effect in a magnonic crystal. The calculation of equal time second-order correlation function showing the quantum blockade effect is given in \ref{blockade}.

%\st{In Fig.}~\ref{QI_DIODE_OTOC}, \st{$C_L(t)$ and $C_R(t)$ is shown for
%$|n^+-m|$ and $|n^--m|$ }
%\st{distant spins, respectively. $C_R(t)$ is
%independent of the separation between the spins, however, the decay
%amplitude varies due to the suppression coefficient $\zeta$. In the
%case of the dominant attenuation by the gate magnons, the OTOC
%decreases significantly. The difference in $C_L$ and $C_R$ is
%originated due to the asymmetry arising from the DMI term.}
 Fig.~\ref{QI_DIODE_OTOC}(a) and Fig.~\ref{QI_DIODE_OTOC}(b) are the variation of $C_L(t)$ and $C_R(t)$ for $|n^+-m|$ and $|n^--m|$ distant spins, respectively. Both show similar behavior with increasing separation between the spins. However, the amplitude of $C_R(t)$ is less than $C_L(t)$ because the decay amplitude of the  $C_R(t)$ varies due to the suppression coefficient $\zeta$. In the case of the dominant attenuation by the gate magnons, the OTOC decreases significantly. The difference in $C_L(t)$ and $C_R(t)$ originated due to the asymmetry arising from the DMI term. The time required to deviate the OTOC from zero increases as the separation between the spins increases. This observation indicates that quantum information flow has a finite "butterfly velocity." On the contrary, the amplitude of OTOC decreases as the separation between the spins increases because the initial amount of quantum information spreads among more spins. At the large time, OTOC again becomes zero because it spreads over the whole system.  Fig.~\ref{QI_DIODE_OTOC}(c) is the behavior of $C_R(t)$ with decreasing suppression coefficient $\zeta$ and fixed value of distance between the spins $r_{1,2}$. As suppression coefficient $\zeta$ decreases, the amplitude of $C_R(t)$ decreases which is an indicator of increasing rectification. A detailed discussion of rectification is in the next subsection.
\par
A high density of magnons can invalidate the assumption of a pure
state or spin-wave approximation that works only for a low density of
magnons. However, the key point in our case is that one has to
distinguish between two sorts of magnons, gate magnons and propagating
nonreciprocal magnons. The density of the propagating magnons can be
regulated in the experiment through a microwave antenna, and one can
always ensure that their density is low enough. It is easy to regulate
the density of the gate magnons, and an experimentally accessible
method is discussed in Ref.~\cite{Chumak2}.
In the magnonic systems, the Kerr nonlinearity may lead to interesting effects, for example, the magnon-magnon entanglement and frequency shift \cite{zhang2019quantum,moslehi2022photon}. On the other hand, we note that DMI term and strong magneto-electric coupling may be responsible for nonlinear coupling terms similar to the magnon Kerr effect. This effect is studied in Ref.~\cite{toklikishvili2023electrically}. 

\subsection{Rectification}
The efficiency of the quantum information diode is given by the rectification coefficient {\it i.e.,} the ratio between the left and right propagating magnons that is calculated by left and right OTOC.  DMI term and non-reciprocal magnon dispersion relations influence the rectification coefficient in two ways. a) directly meaning that in the left and right OTOCs appear different left and right dispersion relations and b) non-directly meaning that magnonic crystal due to the scattering on the gate magnons bans propagation of the drain magnons in one direction only (damping of the OTOC current). This non-reciprocal damping effect is experimentally observed in magnonic crystals \cite{Chumak2}. The non-reciprocal damping enhances the rectification effect and it was not studied in the context of quantum information and OTOC before.
\par
\begin{figure}
  \centering
  \includegraphics[width=0.35\textwidth]{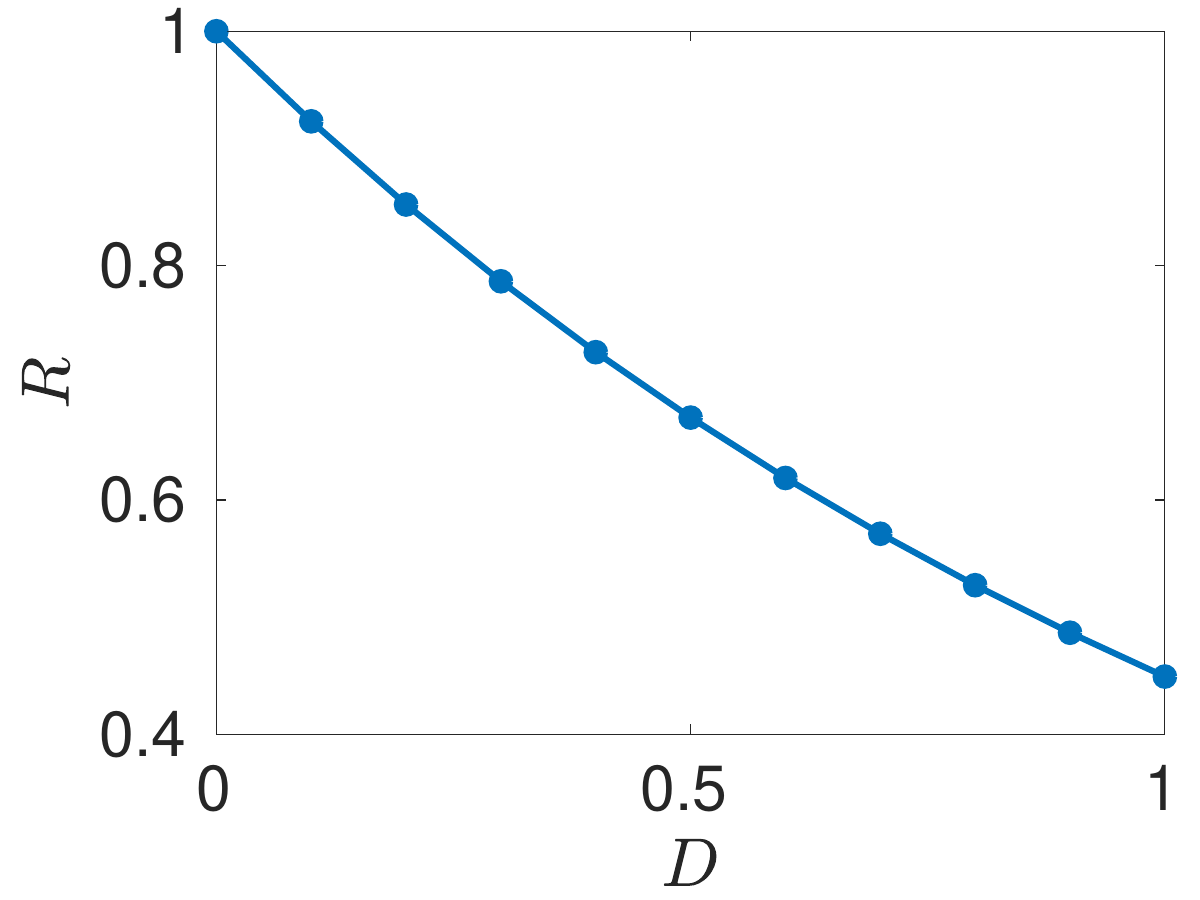}
  \caption{Rectification coefficient $R$ is plotted against DMI coefficient ($D$) for suppression rate $\zeta(D)\approx e^{-D/5}$. The parameters are $J_1=2J_2=1$, $N=1000$,  $r_{12}=10a$, $a_0=1$ and $m_0=1$ to $N$.}
  \label{R}
\end{figure}

Let us calculate the total amount of correlations transferred in opposite directions followed by the rectification coefficient, a function of the external electric field as $R=\frac{\int\limits_0^\infty C_R(t) dt}{\int\limits_0^\infty C_L(t)dt}$.  We interpolate the suppression rate as a function of the DMI coefficient in the form $\zeta(D)\approx e^{-D/5}$.  The coefficient $\zeta(D)$ mimics a scattering process of the drain magnons on the gate magnons \cite{Chumak2}. In Fig.~\ref{R} we see the variation of the rectification coefficient as a function of $D$.  The electric field has a direct and important role in rectification. In particular, DMI constant $D$ depends on the electric field $E_y$ as $D=E_yg_{ME}$, where $g_{ME}$ is the magneto-electric coupling constant. In the case of zero electric field, $D$ will be zero, implying the absence of rectification effect $R=1$. As the electric field increases, $D$ also increases linearly, and rectification decreases exponentially. A detailed study of the role of the electric field in DM has been done in Ref. \cite{Vignale}.

\section{Discussions}
We studied a quantum information flow in a spin quantum system. In
particular, we proposed a quantum magnon diode based on YIG and
magnonic crystal properties. The flow of magnons with wavelengths
satisfying the Bragg conditions $k=m_0\pi/a_o$ is reflected from the
gate magnons. Due to the absence of inversion symmetry in the system, left
and right-propagating magnons have different dispersion relations and
wave vectors. While for the right propagating magnons, the Bragg
conditions hold, left magnons violate them, leading to an asymmetric
flow of the quantum information.
%Fig.\ref{QI_DIODE_OTOC}.
We found that the strength of quantum correlations depends on the
distance between spins and time. The OTOC for the spins separated by
longer distance shows an inevitable delay in time, meaning that the
quantum information flow has a finite "butterfly velocity." On the
other hand, the OTOC amplitude becomes smaller at longer distances
between spins. The reason is that the initial amount of quantum
information spreads among more spins. After the quantum information
spreads over the whole system, which is pretty large ( $N=1000$
sites), the OTOC again becomes zero. 

We proposed a novel theoretical concept that can be directly realized
with the experimentally feasible setup and particular material. There
are several experimentally feasible protocols for measuring OTOC in
the spin systems \cite{nie2020experimental,li2017measuring}. According
to these protocols, one needs to initialize the system into the fully
polarized state, then apply quench and measure the expectation value
of the first spin. All these steps are directly applicable to our
setup from YIG. The fully polarized initial state can be obtained by
switching on and off a strong magnetic field at a time moment t = 0.
Quench, in our case, is performed by a microwave antenna which is an
experimentally accessible device. Polarization of the initial spin can
be measured through the STM tip. Overall our setup is the
experimentally feasible setup studied in Ref.  \cite{Chumak2}.

%\section*{Methods}
%We diagonalize the Hamiltonian of a 2D square-lattice spin system with nearest-neighbor interaction strength $J_1$, next nearest-nearest neighbor interaction strength $J_2$ and DMI interaction strength $D$, given in Eq.~(\ref{Hamiltonian}). In the diagonalization procedure, first, we transform the spins of the Hamiltonian into the bosonic creation ($\hat a^{\dagger}_{n}$) and annihilation operators ($\hat a_{n}$). Afterward, we change the position-dependent annihilation and creation operators into momentum space by using Fourier transformation. A simple algebra leads to the diagonalized Hamiltonian. This diagonalized Hamiltonian helps to calculate the time-evolved operator $\hat \eta_m(t)=\exp(\im \hat H t) \hat \eta_m \exp(-i \hat H t)$, where $\hat\eta_m=1-2\hat a^{\dagger}_m \hat a_m$.
%\par
%In the manuscript, we analytically calculate the formula of left and right OTOC (Detailed calculation is given in the supplementary material). In the calculation of OTOC [$C(t)=1-F(t)$], it is necessary to calculate $F(t)$ that is defined as $F(t)=\langle \hat \eta_m(t)\hat \eta_n\hat \eta_m(t)\hat \eta_n\rangle$. For the calculation of $F(t)$, we take the product of all four terms  $\hat \eta_m(t)$, $\hat \eta_n$, $\hat \eta_m(t)$, and $\hat \eta_n$ and calculate the expectation value over one magnon excitation state. After doing some simple calculations, we get the formula of left and right OTOC.  A detailed derivation is given in the supplementary material.

\section*{Data Availability Statement}
The data that support the findings of this study are available within
the article.

\section*{Acknowledgments}
SKM acknowledges the Science and Engineering Research Board, Department of
Science and Technology, India for support under Core Research Grant
CRG/2021/007095. A.E. acknowledges the funding by the
Fonds zur F\"orderung der Wissenschaftlichen Forschung (FWF) under Grant No. I 5384.
\appendix
\section{Diagonalization of Hamiltonian Eq.~(2)}
\label{supp0}
2D square-lattice spin system with nearest-neighbor $J_1$ and the next nearest-neighbor $J_2$ coupling constants (taking $\hslash=1$): 
\begin{eqnarray}
\label{Hamiltonian_app}
\hat{H}&=&J_1\sum\limits_{\langle
 n,m\rangle}\hat\sigma_n\hat\sigma_m+
  J_2\sum\limits_{\langle\langle
  n,m\rangle\rangle}\hat\sigma_n\hat\sigma_m-{\bf
  P}\cdot{\bf E}, \nonumber \\ 
  &=&J_1\sum\limits_{\langle
 n,m\rangle}\hat\sigma_n\hat\sigma_m+
  J_2\sum\limits_{\langle\langle
  n,m\rangle\rangle}\hat\sigma_n\hat\sigma_m-D\sum\limits_{n}(\hat\sigma_{n}\times\hat\sigma_{n+1})_z, \nonumber \\
  &=&4\Big[J_1\sum\limits_{\langle
 n,m\rangle}\hat S_n\hat S_m+
  J_2\sum\limits_{\langle\langle
  n,m\rangle\rangle} \hat S_n\hat S_m+\frac{D}{i}\sum_n( \hat S_{n}^+\hat S_{n+1}^{-}-\hat  S_{n}^{-}\hat S_{n+1}^{+})\Big], \nonumber \\
  &=&4\Big[J_1 \sum\limits_{\langle
 n,m\rangle} \frac{1}{2}\Big\{\Big(\hat S_n^{-}\hat S_m^{+}+\hat S_n^{+}\hat S_m^{-}\Big)+\hat S_n^{z}\hat S_m^{z}\Big\}
  +J_2\sum\limits_{\langle\langle
  n,m\rangle\rangle} \frac{1}{2}\Big\{\Big(\hat S_n^{-}\hat S_m^{+}+\hat S_n^{+}\hat S_m^{-} \Big) \nonumber \\
  &+&\hat S_n^{z}\hat S_m^{z}\Big\}+\frac{D}{i}\sum_n( \hat S_{n}^+\hat S_{n+1}^{-}-\hat  S_{n}^{-}\hat S_{n+1}^{+})\Big].
\end{eqnarray} 

Spin-half systems have two permitted states on each site, {\it i.e.}, $\vert \uparrow\rangle$ and $\vert \downarrow\rangle$. Operation of spin operators on these state are given as
\begin{equation}
    \hat S^+\vert \downarrow\rangle=\vert \uparrow\rangle,~~ \hat S^+\vert \uparrow\rangle=0,
\end{equation}
\begin{equation}
    \hat S^-\vert \uparrow\rangle=\vert \downarrow\rangle,~~ \hat S^-\vert \downarrow\rangle=0,
\end{equation}
\begin{equation}
    \hat S^z\vert \uparrow\rangle=\frac{1}{2}\vert \uparrow\rangle,~~ \hat S^z\vert \downarrow\rangle=-\frac{1}{2}\vert \downarrow\rangle,
\end{equation}
%Hard-core Bosons allows one or zero boson on each sites, as $\vert 1\rangle$ and $\vert 0\rangle$. Effect of creation and annihilation operators are given as
%\begin{eqnarray}
%&&\hat a \vert 1\rangle =\vert 0\rangle,~ \hat a \vert 0\rangle =0, \nonumber \\
%&& \hat a^{\dagger} \vert 1\rangle =0,~ \hat a^{\dagger} \vert 0\rangle =\vert 1 \rangle,
%\end{eqnarray}
%\begin{eqnarray}
%\hat n\vert 1\rangle = \hat a^{\dagger} \hat a \vert 1\rangle =1,~ \hat n\vert 0\rangle =\hat %a^{\dagger} \hat a \vert 0\rangle =0.
%\end{eqnarray}
Transformation of the spin operators in hard-core bosonic creation and annihilation operators are given as
\begin{eqnarray}
&&\hat S_{m,n}^+=\hat a_{m,n}, \nonumber \\
&& \hat S_{m,n}^-=\hat a_{m,n}^{\dagger}, \nonumber \\
&&  \hat S_{m,n}^z=1/2-\hat a_{m,n}^{\dagger}\hat a_{m,n}
\end{eqnarray}
%\begin{eqnarray}
%\hat H_s&=&\frac{1}{4}\Big[J_1\sum\limits_{\langle
% n,m\rangle}\hat S_n\hat S_m+
 % J_2\sum\limits_{\langle\langle
 % n,m\rangle\rangle} \hat S_n\hat S_m+\frac{D}{i}\sum_n( \hat S_{n}^+\hat S_{n+1}^{-}-\hat  S_{n}^{-}\hat S_{n+1}^{+})\Big]. \nonumber \\
  %&=&\frac{1}{4}\Big[J_1\sum\limits_{\langle
 %n,m\rangle}\hat S_n\hat S_m+
  %J_2\sum\limits_{\langle\langle
  %n,m\rangle\rangle} \hat S_n\hat S_m+\frac{D}{i}\sum_n( \hat S_{n}^+\hat S_{n+1}^{-}-\hat  S_{n}^{-}\hat S_{n+1}^{+})\Big].
%\end{eqnarray} 
Hamiltonian in the bosonic representation is given as 
\begin{eqnarray}
\hat H &=&2\Big[J_1 \sum\limits_{\langle
 n,m\rangle} \Big(\hat a_n^{\dagger}\hat a_m+\hat a_n\hat a_m^{\dagger}-\hat a_n^{\dagger}\hat a_n-\hat a_m^{\dagger}\hat a_m+\frac{1}{2}+\hat a_n^{\dagger}\hat a_n\hat a_m^{\dagger}\hat a_m\Big) \nonumber \\
  &+&J_2\sum\limits_{\langle\langle
  n,m\rangle\rangle} \Big(\hat a_n^{\dagger}\hat a_m+\hat a_n\hat a_m^{\dagger}-\hat a_n^{\dagger}\hat a_n-\hat a_m^{\dagger}\hat a_m+\frac{1}{2}+\hat a_n^{\dagger}\hat a_n\hat a_m^{\dagger}\hat a_m\Big)
  \nonumber \\
  &+&\frac{D}{i}\sum_n \Big(\hat a_{n}\hat a_{n+1}^{\dagger}-\hat a_{n}^{\dagger}\hat a_{n+1}\Big)\Big].\nonumber \\
\end{eqnarray}
Fourier transform of $\hat a_{n}^{\dagger}(\hat a_{n})$ is $\hat a_{\vec{k}}^{\dagger}(\hat a_{\vec{k}})$.
\begin{eqnarray}
&&\hat a_{\vec{k}}^{\dagger}=\frac{1}{\sqrt{N}}\sum_ne^{i \vec{k}\vec{r}_n}a_n^{\dagger},
\nonumber \\ 
&&\hat a_{\vec{k}}=\frac{1}{\sqrt{N}}\sum_ne^{-i \vec{k}\vec{r}_n}a_n.
\end{eqnarray}
Inverse Fourier transform is given as
\begin{eqnarray}
&&\hat a_n^{\dagger}=\frac{1}{\sqrt{N}}\sum_ne^{i \vec{k}\vec{r}_n}a_{\vec{k}}^{\dagger},
\nonumber \\ 
&&\hat a_n=\frac{1}{\sqrt{N}}\sum_ne^{-i \vec{k}\vec{r}_n}a_{\vec{k}}.
\end{eqnarray}

After summing over $n$ we get Hamiltonian  (Eq.~\ref{Hamiltonian_app}) in $\vec{k}$ space as 
%\begin{equation}
%\hat H_s= \sum\limits_{\vec{k}}\omega_{\vec{k}} \hat{a}^{\dagger}_{\vec{k}}\hat{a}_{\vec{k}} +J_1\sum_{n,m}(\hat n_{n,m}\hat n_{n\pm1,m}+\hat n_{n,m}\hat n_{n,m\pm 1})+J_2\sum_{n,m}\hat n_{n,m}\hat n_{n\pm 1,m\pm1} -D\sum_{\vec{k}} \sin(\vec{k} a)\hat a_{\vec{k}}^{\dagger} \hat a_{\vec{k}}.%-D\sum\limits_{n}(\hat\sigma_{n}\times\hat\sigma_{n+1})_z.
%\end{equation}
%Since this is a FM system, the single spin-flip on top of the fully polarized FM ground state can be exactly mapped to a single hard-core boson in the vacuum state. 
%As it known to all, in FM case, the Hamiltonian ignoring the interacting terms is still exact for single-magnon sectors. Hence, we can substitute the effective Hamiltonian
\begin{eqnarray}
\hat H &=& \sum\limits_{\vec{k}}\omega_{\vec{k}} \hat{a}^{\dagger}_{\vec{k}}\hat{a}_{\vec{k}}-D\sum_{\vec{k}} \sin(\vec{k} a)\hat a_{\vec{k}}^{\dagger} \hat a_{\vec{k}}, \nonumber \\
&=&\sum\limits_{\vec{k}}\omega(\pm D,\textbf{k})\hat{a}^{\dagger}_{\vec{k}}\hat{a}_{\vec{k}}
\end{eqnarray}
where,
 \begin{eqnarray}%\label{2Hamiltonian}
 &&\omega(\pm D,\textbf{k})=\big(\omega(\vec{k})\pm\omega_{DM}(\vec{k})\big),  ~~\omega_{DM}(\vec{k})=D\sin(k_xa),\nonumber\\
 &&\omega(\vec{k})=2J_1(1-\gamma_{1,\textbf{k}})+2J_2(1-\gamma_{2,\textbf{k}}),\,\,\gamma_{1,\textbf{k}}=\frac{1}{2}(\cos k_xa+\cos k_ya),\nonumber\\
 && \gamma_{2,\textbf{k}}=\frac{1}{2}[\cos (k_x+k_y)a+\cos (k_x-k_y)a].
  \end{eqnarray}

%%%%%%%%%%%%%%%%%%%%%
%DM interaction Hamiltonian is given as:
%\begin{eqnarray}
%\label{DM}
 %   \hat H_{DM}&=&-D\sum\limits_{n}(\hat\sigma_{n}\times\hat\sigma_{n+1})_z, \nonumber \\
 %   &=& \frac{D}{4i}\sum_n( \hat S_{n}^+\hat S_{n+1}^{-}-\hat  S_{n}^{-}\hat S_{n+1}^{+})
%end{eqnarray}
%Since, $\hat S_{m,n}^+=\sqrt{2}\hat a_{m,n}$
%and $\hat S_{m,n}^-=\sqrt{2}\hat a_{m,n}^{\dagger}$ then, Eq.~(\ref{DM}) can be written as 
%\begin{equation}
%\hat H_{DM}= \frac{D}{2i}\sum_n(\hat a_{n}\hat a_{n+1}^{\dagger}-\hat a_{n}^{\dagger}\hat a_{n+1}),
%\end{equation}
%Fourier transform of $\hat a_{n}^{\dagger}(\hat a_{n})$ is $\hat a_{k}^{\dagger}(\hat a_{k})$. After summing over $n$ we get
%\begin{eqnarray}
%\hat H_{DM} &=& \frac{D}{2i} \sum_k (e^{-i k a}\hat a_{k}\hat a_{k}^{\dagger}-e^{i k a}\hat a_{k}^{\dagger}\hat a_{k}), \nonumber \\
%&=&-D\sum_k \sin(ka)\hat a_{k}^{\dagger} \hat a_{k}.
%\end{eqnarray}

\section{Calculation of left and right out-of-time ordered correlation functions}
\label{supp1}
We will calculate OTOC exactly for one magnon excitation state given in Eq. (7) as

\begin{eqnarray}
\label{occupation number_app}
C(t)=\frac{1}{2}\bigg\lbrace\langle\hat \eta_n\hat \eta_m(t)\hat \eta_m(t)\hat \eta_n\rangle
&+&\langle \hat \eta_m(t)\hat \eta_n\hat \eta_n\hat \eta_m(t)\rangle \nonumber \\
      &-&\langle \hat \eta_m(t)\hat \eta_n\hat \eta_m(t)\hat \eta_n\rangle-\langle
          \hat \eta_n\hat \eta_m(t)\hat \eta_n\hat \eta_m(t)\bigg\rbrace.
         \end{eqnarray}
Here, $\hat \eta_{m/n}= \hat \sigma_{m/n}^z$ is Hermitian and unitary, therefore, Eq.~(\ref{occupation number_app}) transforms in the form given as
\begin{eqnarray}
C(t)=1-\langle \hat \eta_m(t)\hat \eta_n\hat \eta_m(t)\hat \eta_n\rangle=1-F(t),
%=1-\langle \hat \sigma_m(t)\hat \sigma_n\hat \sigma_m(t)\hat \sigma_n\rangle=1-F(t),
\end{eqnarray}
where $F(t)$ is given as
\begin{eqnarray}
\label{F_t}
F(t)&=&\langle \phi \vert \hat a_{n}\hat \eta_m(t)\hat \eta_n\hat \eta_m(t)\hat \eta_na^{\dagger}_
{n}\vert \phi \rangle.
\end{eqnarray}

 In the above equation, the expectation value is taken over one magnon excitation state $\hat a^{\dagger}_n\vert \phi\rangle$, where $\vert \phi\rangle$ is the vacuum state, equivalent to a polarized state. 
First of all we calculate the product of four observables in $F(t)$ (Eq.~(\ref{F_t}))in bosonic representation as
\begin{eqnarray}
\label{S3}
\hat \eta_m(t)\hat \eta_n\hat \eta_m(t)\hat \eta_n&=&[1-2\hat a^{\dagger}_{
m}\hat a_{
m}(t)][1-2\hat a^{\dagger}_{
n}\hat a_{
n}][1-2\hat a^{\dagger}_{
m}\hat a_{
m}(t)][1-2\hat a^{\dagger}_{
n}\hat a_{
n}], \nonumber \\
&=&\Big[1-2\hat a^{\dagger}_{
m}\hat a_{
m}(t)-2\hat a^{\dagger}_{
n}\hat a_{
n}+4\hat a^{\dagger}_{
m}\hat a_{
m}(t)\hat a^{\dagger}_{
n}\hat a_{
n}\Big] \nonumber \\
&\times&\Big[1-2\hat a^{\dagger}_{
m}\hat a_{
m}(t)-2\hat a^{\dagger}_{
n}\hat a_{
n}+4\hat a^{\dagger}_{
m}\hat a_{
m}(t)\hat a^{\dagger}_{
n}\hat a_{
n}\Big], \nonumber \\
&=&1-4\hat a^{\dagger}_{
m}\hat a_{
m}(t)-4\hat a^{\dagger}_{
n}\hat a_{
n}+4\hat a^{\dagger}_{
m}\hat a_{
m}(t)\hat a^{\dagger}_{
n}\hat a_{
n} +4\hat a^{\dagger}_{
n}\hat a_{
n}\hat a^{\dagger}_{
m}\hat a_{
m}(t)\nonumber \\ 
&+& 4\hat a^{\dagger}_{
m}\hat a_{
m}\hat a^{\dagger}_{
m}\hat a_{
m}(t)+4\hat a^{\dagger}_{
n}\hat a_{
n}\hat a^{\dagger}_{
n}\hat a_{
n}+4\hat a^{\dagger}_{
m}\hat a_{
m}\hat a^{\dagger}_{
n}\hat a_{
n}+4\hat a^{\dagger}_{
n}\hat a_{
n}\hat a^{\dagger}_{
m}\hat a_{
m} \nonumber \\
&-&8\hat a^{\dagger}_{
m}\hat a_{
m}\hat a^{\dagger}_{
m}\hat a_{
m}(t) \hat a^{\dagger}_{
n}\hat a_{
n}-8\hat a^{\dagger}_{
n}\hat a_{
n}\hat a^{\dagger}_{
m}\hat a_{
m}(t) \hat a^{\dagger}_{
n}\hat a_{
n} \nonumber \\
&-&8\hat a^{\dagger}_{
m}\hat a_{
m}(t)\hat a^{\dagger}_{
n}\hat a_{
n}\hat a^{\dagger}_{
m}\hat a_{
m}(t)-8\hat a^{\dagger}_{
m}\hat a_{
m}(t)\hat a^{\dagger}_{
n}\hat a_{
n} \hat a^{\dagger}_{
n}\hat a_{
n} \nonumber \\
&+& 16\hat a^{\dagger}_{
m}\hat a_{
m}(t)\hat a^{\dagger}_{
n}\hat a_{
n}\hat a^{\dagger}_{
m}\hat a_{
m}(t)\hat a^{\dagger}_{
n}\hat a_{
n}.
\end{eqnarray}
Further, we calculate the expectation value of the last term of  Eq. (\ref{S3}) over one magnon excitation state  {\it i. e.},
%\begin{eqnarray}
%\label{S4}
%&=&
$\langle \phi \vert \hat a_{n} \hat a^{\dagger}_{
m}\hat a_{
m}(t) \hat a^{\dagger}_{n}\hat a_{n}
\hat a_{n} \hat a^{\dagger}_{
m}\hat a_{
m}(t) \hat a^{\dagger}_{n}\hat a_{n}\hat a^{\dagger}_{n}\vert \phi \rangle, $
%\end{eqnarray}
using the properties of bosonic operators $[\hat a_i, \hat a_j^{\dagger}]=\delta_{ij}$, $(\hat a_i)^2=0$, and $(\hat a^{\dagger}_i)^2=0$. We get 
\begin{eqnarray}
\label{S5}
\langle \phi \vert \hat a_{n} \hat a^{\dagger}_{m}\hat a_{m}(t) \hat a^{\dagger}_{n}\hat a_{n}\hat a_{n} \hat a^{\dagger}_{m}\hat a_{m}(t) \hat a^{\dagger}_{n}\hat a_{n}\hat a^{\dagger}_{n}\vert \phi \rangle
&=\langle \phi \vert \hat a_{n} e^{i\hat H t}\hat a^{\dagger}_{
m}\hat a_{
m}e^{-i\hat H t} \hat a^{\dagger}_{n}\hat a_{n}
e^{i\hat H t}\hat a^{\dagger}_{
m}\hat a_{
m}e^{-i\hat H t} \hat a^{\dagger}_{n}\vert \phi \rangle,& \nonumber \\
&= \langle \Psi(t)\vert \Psi(t)  \rangle,
\end{eqnarray}
where $\vert \Psi(t)\rangle=\hat a_{n}
e^{i\hat H t}\hat a^{\dagger}_{
m}\hat a_{
m}e^{-i\hat H t} \hat a^{\dagger}_{n}\vert \phi \rangle.$ Fourier transformation of the $\vert \Psi(t)\rangle$ and  diagonalized Hamiltonian will provide
\begin{eqnarray}
\vert \Psi(t)\rangle &=&\frac{1}{N}\sum_k e^{i(-k(m-n)+\omega_kt/\hslash)} \frac{1}{N}\sum_{k^{'}} e^{i( k^{'} (m-n)-\omega_{k^{'}}t/\hslash)} \vert \phi \rangle  \nonumber \\
&=&\frac{1}{N^2} \Omega_1 \Omega_2 \vert \phi \rangle. \nonumber
\end{eqnarray}
Hence, 
\begin{eqnarray}
\label{S6}
\langle \Psi(t)\vert \Psi(t)  \rangle=\frac{1}{N^4}\Omega_1 \Omega_2\Omega_1 \Omega_2.
\end{eqnarray}
Similarly,
\begin{eqnarray}
\label{S7}
\langle \phi \vert \hat a_{n}
 \hat a_{
m} \hat a_{
m}(t)\hat a^{\dagger}_{n} \vert \phi \rangle=\frac{1}{N^2}\Omega_1 \Omega_2
\end{eqnarray}
After doing simple bosonic algebra, time-dependent terms of Eq.~(\ref{S3}) are converted either in the form of Eq.~(\ref{S5}) or Eq.~(\ref{S7}). By using Eq.~(\ref{S6}) and Eq.~(\ref{S7}), we calculate $F(t)$ as
\begin{eqnarray}
F(t)&=1-\frac{4}{N^2}\Omega_1\Omega_2-4+\frac{4}{N^2}\Omega_1\Omega_2+\frac{4}{N^2}\Omega_1\Omega_2+\frac{4}{N^2}\Omega_1\Omega_2+\frac{4}{N^2}\Omega_1\Omega_2+\frac{4}{N^2}\Omega_1\Omega_2+4 &\nonumber \\
&-\frac{8}{N^2}\Omega_1\Omega_2-\frac{8}{N^2}\Omega_1\Omega_2 -\frac{8}{N^4}\Omega_1\Omega_2\Omega_1\Omega_2-\frac{8}{N^2}\Omega_1\Omega_2+\frac{16}{N^4}\Omega_1\Omega_2\Omega_1\Omega_2 &\nonumber \\ 
&=1-\frac{8}{N^2}\Omega_1\Omega_2+\frac{8}{N^4}\Omega_1\Omega_2\Omega_1\Omega_2 
\end{eqnarray}
Then, we get the left and right OTOCs’ analytical expressions as

\begin{eqnarray}
\label{we deduce the analytical formula2}
C_L(t)&=&\frac{8}{N^2}\Omega_1^L\Omega_2^L-\frac{8}{N^4}\Omega_1^L\Omega_2^L\Omega_1^L\Omega_2^L,  \nonumber \\
C_R(t)&=&\zeta^4(D)\Big(\frac{8}{N^2}\Omega_1^R\Omega_2^R-\frac{8}{N^4}\Omega_1^R\Omega_2^R\Omega_1^R\Omega_2^R\Big),  
\end{eqnarray}
where frequencies $\Omega_{1/2}^{L/R}$ are given as
\begin{eqnarray}\label{Left and Right}
\Omega_1^R&=&\Omega_2^{R*}=\sum_{m_0}\exp\bigg(-\frac{im_0\pi r_{1,2}} {a_0}\bigg)\exp\bigg(\frac{i \omega_{m_0}t}{ \hslash}\bigg),
\nonumber
\end{eqnarray}
and
\begin{eqnarray}
\Omega_1^L&=&\Omega_2^{L*}=\sum_{m_0} \exp(-ik_s^{-}r_{1,2})\exp\bigg(\frac{i \omega_{m_0}t}{ \hslash}\bigg).
\end{eqnarray}
\section{Quantum blockade effects}
\label{blockade}
%In the calculation of $F(t)=\langle \phi \vert \hat a_{n}\hat \eta_m(t)\hat \eta_n\hat \eta_m(t)\hat \eta_na^{\dagger}_
%{n}\vert \phi \rangle$ [Eq.~(\ref{F_t})], the expectation value is take over one magnon excitation state $\hat a^{\dagger}_n\vert \phi\rangle$, where $\vert \phi\rangle$ is the vacuum state, equivalent to a polarized state. 
To analyze the magnon blockade effect, we calculate the equal time second-order correlation function defined as \cite{liu2019magnon,xie2020quantum,wu2021phase,zhao2020simultaneous,wang2022hybrid,wang2020magnon}
\begin{equation}
    \label{blockade_eq}
g_a^{2}(0)=\frac{\text{Tr}(\hat \rho \hat a_m^{\dagger 2}\hat a_m^2)}{\left[\text{Tr}(\hat\rho\hat a_m^{\dagger}\hat a_m)\right]^2}=\frac{\langle\hat a_m^{\dagger 2}\hat a_m^2\rangle}{\langle a_m^{\dagger}\hat a_m\rangle^2},
\end{equation}
where $a_m(a_m^{\dagger})$  are  the  annihilation  (creation)  operators  of  the magnon excitation. The magnon blockade is inferred from the condition $g_a^{2}(0)\rightarrow 0$ meaning that magnons can be excited individually, and two or more magnons cannot be excited together.
\par
We note that
    $ \hat a_m^2\hat a_m^{\dagger}\vert \phi\rangle=0$ leading to
$g_a^2(0)=0$. Therefore, the quantum blockade effect occurs in this case.
\section*{References}
\bibliographystyle{iopart-num}
\bibliography{qidiode_ref}
%\section*{Author contributions}
%All authors contributed equally to this work. L.C., A.E., S.S.P.P. and
%S.K.M. conceived the project. R.K.S., H.V., and V.V. performed the numerical
%and analytical calculations. All authors worked on the interpretation
%of the results and contributed to the writing of the manuscript.

%\section*{Competing interests}
%The authors declare no competing financial or non-financial interests.

%\section*{Additional information}
%See Supplementary Information for diagonalization procedure of
%Hamiltonian Eq.~(\ref{Hamiltonian}) resulting in
%Eq.~(\ref{2Hamiltonian}) and  left $C_L(t)$ and right $C_R(t)$ OTOC
%calculation. 

\end{document}